# A contribution to the history of quarks: Boris Struminsky's 1965 JINR publication

Fyodor V. Tkachov

Institute for Nuclear Research of Russian Academy of Sciences
Moscow 117312
Russian Federation

*Abstract.* The idea that an internal quantum number for quarks, eventually called color, can help explain the magnetic moments of baryons within the framework of standard quantum theory, was explicitly stated in a 7th January, 1965 JINR publication by Boris Struminsky [1]. A complete translation into English of Struminsky's letter is presented along with page images of the original booklet found in the JINR Library.

---

*This piece is not concerned with any issue in regard of the discovery of quark color except the relation between the booklets [1] and [2].*

---

Few would object in public that the research community should strive to ensure that every researcher get their proper share of credit for their effort. It is never too late to straighten out a record if only because it would be nice for some of us to know that our research records do stand a chance of an eventual straightening-out. Also, I am confident that Boris Struminsky's former colleagues and friends should be happy that his contribution is finally made available to the public.

A notion widely echoed within the Russian theoretical physics community is that, along with the famous 1965 publications by M.Y. Han and Y. Nambu, and Y. Miyamoto (O.W. Greenberg ought to be mentioned too, I think), the quark quantum number subsequently called color was also independently discovered in the publication [2] by Nikolai Bogoliubov (1909-1992), Boris Struminsky (1939-2003) and Albert Tavkhelidze (born 1930). Since Mr. Tavkhelidze has been, for a very long time already, brandishing a considerable administrative influence within the Russian theoretical physics community, it is not surprising that the notion — with explicit hints at the Nobel Prize quality of Mr. Tavkhelidze's deed — is being actively promulgated by his trustful and loyal associates (cf. the recent flurry of interviews [5]).

Having been in a position to observe Mr. Tavkhelidze since 1978 in various brilliant capacities — a lecturer, a scientific advisor (he was in fact listed as my Ph.D. advisor in 1979-1983), the (still current) head of a theoretical physics department, and, for a number of years (1970-1986), the director of my home research institute, — I felt ashamed that, in view of my somewhat mathematical and computational inclinations, I was not properly familiar with that particular episode of the history of quarks. The reading of the original research accounts invariably proving an uplifting experience, but even the KEK KISS Library not having scans of the early preprints issued at the Joint Institute for Nuclear Research in Dubna, I figured the JINR Library surely must have an archived paper copy of a JINR publication. It proved indeed to be the case as I discovered on 17th December, 2008 during a scientific visit to JINR, where I used spare forty minutes between lunch and a report to a committee to visit the JINR Library on the ground floor, near the entrance, in the Bogoliubov Laboratory of Theoretical Physics, where every JINR theorist and visitor has an unencumbered access.

If one actually reads ref. [2], one cannot fail to miss an explicit reference made there to an earlier publication [1], in regard of an additional quantum number being necessary within the framework of standard quantum theory for the quark theory to agree with experimental data on magnetic moments of baryons. Enchanted by the heroic atmosphere of the quark theory in the making, I swiftly turned back to the catalogue where, somewhat strangely, there were no catalogue cards for Struminsky, for which the lady on duty, somewhat surprised, could provide no explanation. I subsequently learned that Struminsky's publications are properly represented in the main catalogue in a back room, but I was dealing with the one in the front room immediately accessible to all visitors. Anyway, the conscientious lady promised that the publication I was seeking (its JINR number was listed in ref. [2]) must surely be available in the reading hall down the corridor. Where, indeed, another helpful lady brought me the requested material from another back room. The new quark internal degree of freedom was explicitly mentioned in a footnote on page 3 of the booklet (the first page of the text proper, see below), and I realized that the work deserved to be examined without haste. Fortunately, my Sony T200 camera produced adequate snapshots even without proper lighting that I had absolutely no time to arrange. The original camera photos are stored on the web page [6], and included below are reduced black-and-white images.

It should perhaps be clarified that any publication activity in the Union of Soviet Socialist Republics was regarded as a potential means of the class-enemy propaganda — i.e. as an activity highly suspect by default and therefore tightly controlled. Apparently, the only way the concept of preprint could be made compatible with the System was by treating it, essentially, as a book — implying an amount of bureaucratic motions to go through, a registration with the Union libraries and depositories, and an obligatory thick cover. Adding to that a whopping 400+ number of copies produced for each JINR preprint (to be precise, 440 copies indicated on the back cover of [1]), along with the fact that the international institution sent them out to scientific libraries all around the globe, it was apparently deemed adequate and sufficient, even with the on-line arXivs non-existent, to limit the research publication effort to issuing such a booklet, regarded rather like an issue of a journal. This may explain why, for a period of time, some papers published at JINR in the form of such booklets were not submitted to regular journals.

A related point closer to our present subject, is that such a publication could not have occurred without obtaining all the necessary formal permissions, approvals and signatures, including one of the director of the institution or at least the head of the theory department. This means that there is no way ref. [1] could have seen the light of day without explicit approval and permission by Bogoliubov. Indeed, Struminsky was Bogoliubov's Ph.D. student but registered in the Moscow State (Lomonosov) University rather than JINR [7]. Ref. [1] was followed by three more publications in the same year [2]-[4], which demonstrate Struminsky's scientific activity at the time.

Another point worthy of notice is an obvious parallel with the famous publications by Bardeen, Cooper and Schrieffer. A thorough historical account [9] records that Bogoliubov developed his microscopic theory of superconductivity (the one based on what entered textbooks as the Bogoliubov transformation) between Cooper's Phys. Rev. Letter and the main BCS publication in 1958. The emotional interactions between Bogoliubov and Landau in connection with that theory, as described in [9], leave no room for doubt that the story of Cooper's letter preceding the main publication and thus ensuring, in effect, that a member of the collaboration gets a proper credit for his breakthrough contribution to the joint effort, was burnt into Bogoliubov's memory. Struminsky's publication [1] precedes a more complete treatment [2] in a manner too obviously similar to Cooper's letter, to doubt its intention.

Mr. Tavkhelidze's own account of the discovery, composed years after both his collaborators' demise, is somewhat abstract [10]: *"There emerged an idea that quarks have an additional quantum number — color ..."* It is true that ref. [2] made an important contribution to clarifying the dynamic aspect of the quark magnetic moments (the role played by quarks' energy rather than mass; for more on this see the noteworthy review [11]) — but the symmetry argument leading to the additional quantum number was fully and explicitly present in the booklet [1].

From 1968 on, Struminsky published from the Institute of Theoretical Physics in Kiev, Ukraine, where he resided till his demise in 2003. Boris Vladimirovich attended a conference in mid-90's where, unfortunately, I missed the chance to introduce myself to him. But I overheard enough of his remarks to form an impression of him as a sharp, observant and sarcastic person, although his health problem was apparent too. The full significance of one of his remarks in particular, addressed to a close associate of Mr. Tavkhelidze, was not quite clear to me at the time; now, I marvel Struminsky's reserve. And one cannot help wondering whether Struminsky's health condition had not been expedited by his early career experiences.

Boris Vladimirovich published a number of scientific papers, several solo, some of which are listed in the SPIRES database. Mr. Tavkhelidze's name, on the other hand, is listed, with regalia and a photo, on page 751 of a Mathematical Dictionary [8], in company with Student, Tarski and Tartaglia, among other mathematicians (see [6] for a scanned image); but after thirty years under Mr. Tavkhelidze, I cannot, for the life of me, imagine a mathematical discovery he could have possibly made for such an honor to be deserved.

A photo of Dr. Struminsky can be found on the web page [6].

The English translation of ref. [1] is presented below after the references. Then follow page images of the copy of [1] from the JINR Library. The originals can be found on the web page [6].

*Acknowledgements.* I thank my numerous experimental and theoretical colleagues for their acute interest in, and help with this investigation — before and after the posting.

I thank a colleague for pointing out a number of misspellings.

*Special thanks go to the JINR veterans for their supportive responses.*

# Translation of the preprint

P-1939

B.V. Struminsky

## MAGNETIC MOMENTS OF BARYONS IN THE QUARK MODEL

An interesting attempt to unify internal and space symmetries of elementary particles is made in the papers by Gursey, Pais and Radicatti/1/. When the Lorenz group is extended by the SU(3) group, the group $g_6$ emerges whose small subgroup is SU(6). The group SU(6) considered in those papers is a natural generalization of Wigner's group SU(4).

Using the group SU(6), it proves possible to infer some interesting results. In particular, if we put the decuplet of baryons $\left(\frac{3}{2}\right)^+$ and the octet $\left(\frac{1}{2}\right)^+$ into the 56-dimensional representation of SU(6), then the following relation between magnetic moments of baryons is obtained: $\mu_\Omega : \mu_p : \mu_n : \mu_{\Sigma^-} = -3:3:-2:-1$ . (Magnetic moments of the remaining octet components follow from the relations of unitary symmetry

$$\mu_p = \mu_{\Sigma^+},\ \mu_n = \mu_{\Xi^0},\ \mu_{\Sigma^-} = -\mu_{\Sigma^{-\prime}},\ \mu_\Lambda = -\mu_{\Sigma^0} = \tfrac{1}{2}\mu_n).$$

In the quark model/2/ baryons are viewed as bound states of three particles (quarks):

$$u\left(T_3 = \tfrac{1}{2},\ Y = \frac{1}{3},\ Q = \frac{2}{3}\right),\ d\left(T_3 = -\tfrac{1}{2},\ Y = \frac{1}{3},\ Q = -\frac{1}{3}\right),\ s\left(T_3 = 0,\ Y = -\frac{2}{3},\ Q = -\frac{1}{3}\right),$$

The ratio of magnetic moments of quarks: $\mu_u : \mu_d : \mu_s = 2:-1:-1$.

Magnetic moments of baryons can be evaluated using the known formulae of quantum mechanics.

We shall assume that baryons are formed with the three particles in the S-state. Then the $\Omega^-$ hyperon that has the spin $\frac{3}{2}$ and belongs to the unitary decouplet must have a fully antisymmetric space wave function and its magnetic moment is $3\mu_s$ . [**Footnote:** Three identical quarks cannot form an antisymmetric S-state. In order to realize an antisymmetric orbital S-state, it is necessary for the quark to have an additional quantum number.]

For the octet of baryons with spin $\frac{1}{2}$ three symmetry types of the orbital wave function are possible: fully antisymmetric, symmetric and of a mixed symmetry. The complete wave function has the form:

$$\Psi = \Psi_a(x_1)\xi_s + \Psi_s(x_1)\xi_a + \Psi'(x_1)\xi'' - \Psi''(x_1)\xi',$$

where  are functions in the space of spin and unitary spin. They are constructed from functions in the unitary and spin space:

$$\xi_s = \frac{1}{\sqrt{2}}(\kappa' B'' + \kappa'' B')$$

$$\xi_a = \frac{1}{\sqrt{2}}(\kappa' B'' - \kappa'' B')$$

$$\xi' = \frac{1}{\sqrt{2}}\left(\kappa' B'' + \kappa'' B'\right)$$

$$\xi'' = \frac{1}{\sqrt{2}}\left(\kappa' B' - \kappa'' B''\right)$$

$\kappa'$, $\kappa''$ are functions in the spin space:

$$\kappa' = \frac{1}{\sqrt{6}}\left(2b_1a_2a_3 - a_1b_2a_3 - a_1a_2b_3\right)$$

$$\kappa'' = \frac{1}{\sqrt{2}} a_1\left(a_2b_3 - b_2a_3\right),$$

where $a_i$ is the state of the i-th quark with the spin projection $+\frac{1}{2}$, $b_i$ is the state with the spin projection $-\frac{1}{2}$.

Functions in the unitary space have the form:

$$P' = \frac{1}{\sqrt{6}}\left(2d_1u_2u_3 - u_1d_2u_3 - u_1u_2d_3\right), \quad P'' = \frac{1}{\sqrt{2}} u_1\left(u_2d_3 - d_2u_3\right),$$

$$N' = \frac{1}{\sqrt{6}}\left(2u_1d_2d_3 - d_1d_2u_3 - d_1u_2d_3\right), \quad N'' = \frac{1}{\sqrt{2}} d_1\left(d_2u_3 - u_2d_3\right),$$

$$\Xi^{-\prime} = \frac{1}{\sqrt{6}}\left(2d_1s_2s_3 - s_1s_2d_3 - s_1d_2s_3\right), \quad \Xi^{-\prime\prime} = \frac{1}{\sqrt{2}} s_1\left(s_2d_3 - d_2s_3\right),$$

$$\Xi^{0\prime} = \frac{1}{\sqrt{6}}\left(2u_1s_2s_3 - s_1s_2u_3 - s_1u_2s_3\right), \quad \Xi^{0\prime\prime} = \frac{1}{\sqrt{2}} s_1\left(s_2u_3 - u_2s_3\right),$$

$$\Sigma^{+\prime} = \frac{1}{\sqrt{6}}\left(2s_1u_2u_3 - u_1u_2s_3 - u_1s_2u_3\right), \quad \Sigma^{+\prime\prime} = \frac{1}{\sqrt{2}} u_1\left(u_2s_3 - s_2u_3\right),$$

$$\Sigma^{-\prime} = \frac{1}{\sqrt{6}}\left(2s_1d_2d_3 - d_1d_2s_3 - d_1s_2d_3\right), \quad \Sigma^{-\prime\prime} = \frac{1}{\sqrt{2}} d_1\left(d_2s_3 - s_2d_3\right),$$

$$\Sigma^{0\prime} = \frac{1}{2\sqrt{3}}\left\{2s_1\left(u_2d_3 + d_2u_3\right) - s_2\left(u_1d_3 + d_1u_3\right) - s_3\left(u_2d_1 + d_2u_1\right)\right\},$$

$$\Sigma^{0\prime\prime} = \frac{1}{2}\left\{s_3\left(u_2d_1 + d_2u_1\right) - s_2\left(u_1d_3 + d_1u_3\right)\right\},$$

$$\Lambda' = \frac{1}{2}\left\{s_3\left(u_2d_1 - d_2u_1\right) - s_2\left(u_1d_3 - d_1u_3\right)\right\},$$

$$\Lambda'' = \frac{1}{2\sqrt{3}}\left\{s_3\left(u_2d_1 - d_2u_1\right) + s_2\left(u_1d_3 - d_1u_3\right) - 2s_1\left(u_3d_2 - d_3u_2\right)\right\}.$$

Computing the matrix elements of the operator of the spin magnetic moment:

$$\vec{\mu} = \sum_{i=1}^{3}\left(\mu_u P_i^u + \mu_d P_i^d + \mu_s P_i^s\right)\vec{\sigma}_i \,,$$

where $P^{u,d,s}$ are operators projection to the states $u_i, d_i, s_i$, $\sigma_i$ are Pauli matrices acting on the i-th particle, we obtain the following ratio of the magnetic moments:

1) for the antisymmetric space function

$$\mu_\Omega : \mu_p : \mu_n : \mu_{\Sigma^-} = -3 : 3 : -2 : -1 \;;$$

2) for the symmetric space function

$$\mu_p : \mu_n : \mu_{\Sigma^-} = -1 : 2 : -1 \ ;$$

1) for the mixed-symmetry space function

$$\mu_p : \mu_n : \mu_{\Sigma^-} = 2 : 0 : -2 \ .$$

We note that a certain representation of the SU(6) group corresponds to each symmetry type of the space function: the representation of dimensionality 56 corresponds to the antisymmetric one; the representation of dimensionality 20, to the symmetric one; and the representation of dimensionality 70 corresponds to the mixed-symmetry function.

The following circumstance is interesting. As is well known, the deviations of magnetic moments of the nuclei $He^3$ and $H^3$ from theoretical values is explained by the exchange magnetic moment that is positive for $H^3$, and negative for $He^3$. In the quark model, proton is analogous to $He^3$, and neutron, to $H^3$, and the deviations of magnetic moments of the proton $\mu_p = 2.79$ and of the neutron $\mu_p = -1.91$ from theoretical values agree with those for $He^3$ and $H^3$.

The author expresses his sincere gratitude to academician N.N. Bogoliubov for suggesting the problem and attention.

# Page photos of ref. [1] from the JINR library

The full-size camera originals can be found at http://www.inr.ac.ru/~ftkachov/struminsky/.

Front cover:

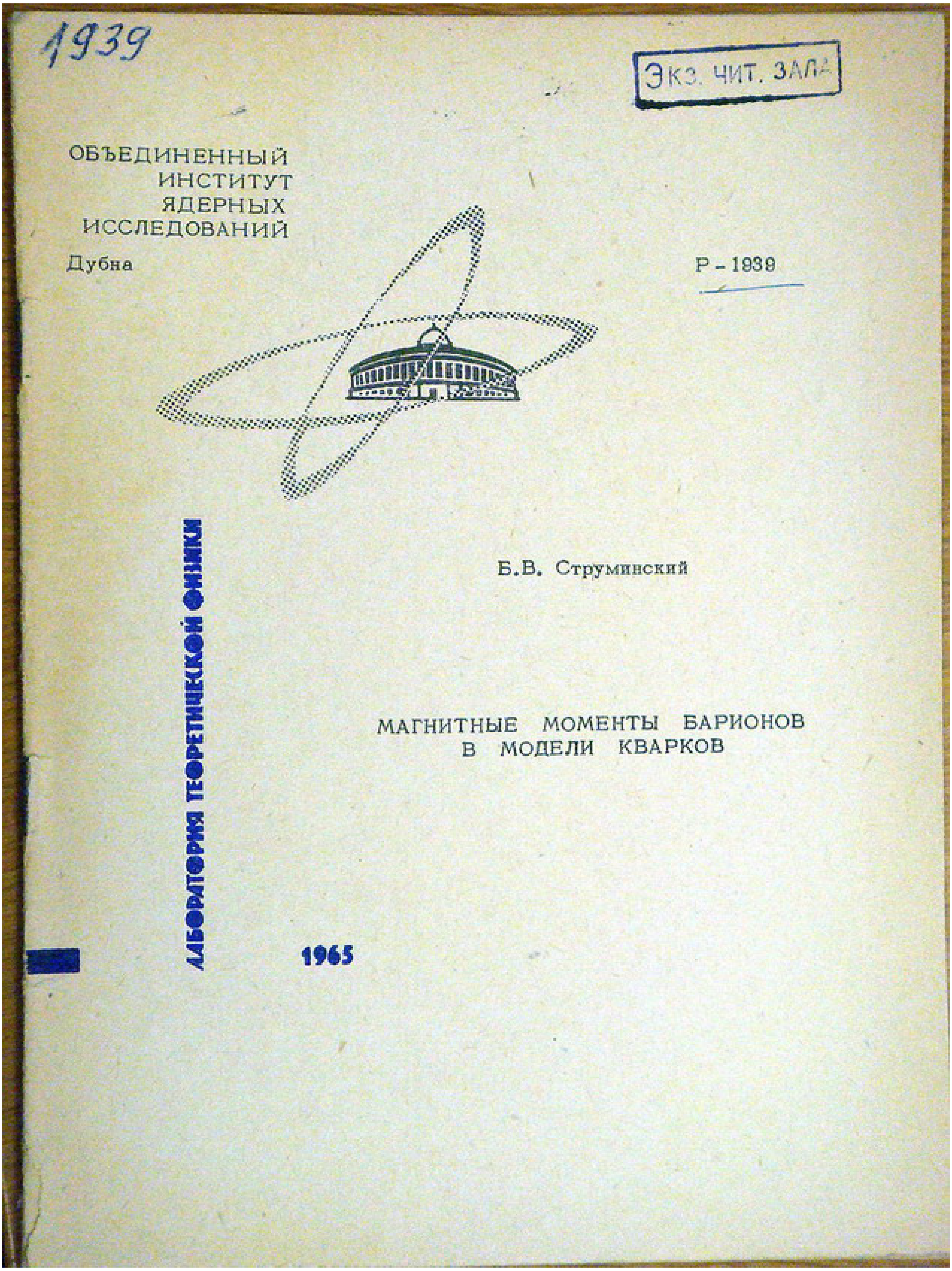

Title page:

Р - 1939

Б.В. Струминский

**МАГНИТНЫЕ МОМЕНТЫ БАРИОНОВ
В МОДЕЛИ КВАРКОВ**

*Stamp at bottom:* ОИЯ И БИБЛИОТЕКА (JINR Library)

Start of text on page 3:

В работах Гюрши, Пайса и Радикатти /1/ сделана интересная попытка объединения внутренней и пространственной симметрии элементарных частиц. При расширении группы Лоренца с помощью группы SU(3) возникает группа $E_6$, малой группой которой является группа SU(6). Группа SU(6), рассматриваемая в этих работах, есть естественное обобщение группы SU(4) Вигнера.

Используя группу SU(6), удается получить ряд интересных следствий. В частности, если мы помещаем декуплет барионов $(\frac{3}{2})^+$ и октет $(½)^+$ в представление размерности 56 группы SU(6), то получается следующее соотношение между магнитными моментами барионов: $\mu_\Omega : \mu_p : \mu_n : \mu_{\Sigma^-} = -3 : 3 : -2 : -1$. (Магнитные моменты остальных компонент октета следуют из соотношений унитарной симметрии

$$\mu_p = \mu_{\Sigma^+},\ \mu_n = \mu_{\Xi^0},\ \mu_{\Sigma^-} = -\mu_{\Sigma^-},\ \mu_\Lambda = -\mu_{\Sigma^0} = ½\mu_n ).$$

В модели кварков /2/ барионы рассматриваются как связанные состояния трех частиц (кварков): $u(T_3 = ½,\ Y = \frac{1}{3},\ Q = \frac{2}{3})$, $d(T_3 = -½,\ Y = \frac{1}{3},\ Q = -\frac{1}{3})$, $s(T_3 = 0,\ Y = -\frac{2}{3},\ Q = -\frac{1}{3})$.

Отношение магнитных моментов кварков: $\mu_u : \mu_d : \mu_s = 2 : -1 : -1$.

Магнитные моменты барионов можно вычислить по известным формулам квантовой механики.

Будем предполагать, что барионы образуются в S -состоянии трех частиц. Тогда $\Omega^-$ гиперон, который имеет спин $\frac{3}{2}$ и принадлежит унитарному декуплету, должен иметь полностью антисимметричную пространственную волновую функцию и его магнитный момент есть $3\mu_s$ x/.

Для октета барионов со спином ½ возможны три типа симметрии орбитальной волновой функции: полностью антисимметричная, симметричная и смешанной симметрии. Полная волновая функция имеет вид:

$$\Psi = \Psi_a(x_i)\ \xi_s + \Psi_s(x_i)\xi_a + \Psi'(x_i)\xi'' - \Psi''(x_i)\xi',$$

где $\xi$ – функции в пространстве спина и унитарного спина. Они строятся из функций в унитарном и спиновом пространстве:

$$\xi_s = \frac{1}{\sqrt{2}}(\kappa' B'' + \kappa'' B')$$

---

x) Три тождественных кварка не могут находиться в антисимметричном S -состоянии. Для того, чтобы реализовать антисимметричное орбитальное S -состояние, надо приписать кварку дополнительное квантовое число.

$$\xi_a = \frac{1}{\sqrt{2}}(\kappa' B'' - \kappa'' B')$$

$$\xi' = \frac{1}{\sqrt{2}}(\kappa' B'' + \kappa'' B')$$

$$\xi'' = \frac{1}{\sqrt{2}}(\kappa' B' - \kappa'' B'').$$

$\kappa'$ , $\kappa''$ функции в спиновом пространстве:

$$\kappa' = \frac{1}{\sqrt{6}}(2b_1 a_2 a_3 - a_1 b_2 a_3 - a_1 a_2 b_3)$$

$$\kappa'' = \frac{1}{\sqrt{2}} a_1 (a_2 b_3 - b_2 a_3),$$

где $a_i$ –состояние $i$ –кварка с проекцией спина $+½$ , $b_i$ –состояние с проекцией спина $-½$ .

Функции в унитарном пространстве имеют вид:

$$P' = \frac{1}{\sqrt{6}}(2d_1 u_2 u_3 - u_1 d_2 u_3 - u_1 u_2 d_3), \quad P'' = \frac{1}{\sqrt{2}} u_1 (u_2 d_3 - d_2 u_3),$$

$$N' = \frac{1}{\sqrt{6}}(2u_1 d_2 d_3 - d_1 d_2 u_3 - d_1 u_2 d_3), \quad N'' = \frac{1}{\sqrt{2}} d_1 (d_2 u_3 - u_2 d_3),$$

$$\Xi^{-\prime} = \frac{1}{\sqrt{6}}(2d_1 s_2 s_3 - s_1 s_2 d_3 - s_1 d_2 s_3), \quad \Xi^{-\prime\prime} = \frac{1}{\sqrt{2}} s_1 (s_2 d_3 - d_2 s_3),$$

$$\Xi^{0\prime} = \frac{1}{\sqrt{6}}(2u_1 s_2 s_3 - s_1 s_2 u_3 - s_1 u_2 s_3), \quad \Xi^{0\prime\prime} = \frac{1}{\sqrt{2}} s_1 (s_2 u_3 - u_2 s_3),$$

$$\Sigma^{+\prime} = \frac{1}{\sqrt{6}}(2s_1 u_2 u_3 - u_1 u_2 s_3 - u_1 s_2 u_3), \quad \Sigma^{+\prime\prime} = \frac{1}{\sqrt{2}} u_1 (u_2 s_3 - s_2 u_3),$$

$$\Sigma^{-\prime} = \frac{1}{\sqrt{6}}(2s_1 d_2 d_3 - d_1 d_2 s_3 - d_1 s_2 d_3), \quad \Sigma^{-\prime\prime} = \frac{1}{\sqrt{2}} d_1 (d_2 s_3 - s_2 d_3),$$

$$\Sigma^{0\prime} = \frac{1}{2\sqrt{3}}\{2s_1(u_2 d_3 + d_2 u_3) - s_2(u_1 d_3 + d_1 u_3) - s_3(u_2 d_1 + d_2 u_1)\}$$

$$\Sigma^{0\prime\prime} = ½\{s_3(u_2 d_1 + d_2 u_1) - s_2(u_1 d_3 + d_1 u_3)\}$$

$$\Lambda' = ½\{s_3(u_2 d_1 - d_2 u_1) - s_2(u_1 d_3 - d_1 u_3)\}$$

$$\Lambda'' = \frac{1}{2\sqrt{3}}\{s_3(u_2 d_1 - d_2 u_1) + s_2(u_1 d_3 - d_1 u_3) - 2s_1(u_3 d_2 - d_3 u_2)\}.$$

Вычислив матричные элементы оператора спинового магнитного момента:

$$\vec{\mu} = \sum_{i=1}^{3} (\mu_u P_i^u + \mu_d P_i^d + \mu_s P_i^s)\vec{\sigma}_i ,$$

где $P^{u,d,s}$ — операторы проектирования на состояния $u_i$ , $d_i$ , $s_i$ , $\sigma_i$ -матрицы Паули, действующие на $i$ -частицу, получим следующее отношение магнитных моментов:

1) для антисимметричной пространственной функции

$$\mu_\Omega : \mu_p : \mu_n : \mu_{\Sigma^-} = -3 : 3 : -2 : -1 ;$$

2) для симметричной пространственной функции

$$\mu_p : \mu_n : \mu_{\Sigma^-} = -1 : 2 : -1 ;$$

3) для пространственной функции смешанной симметрии

$$\mu_p : \mu_n : \mu_{\Sigma^-} = 2 : 0 : -2 .$$

Заметим, что каждому типу симметрии пространственной функции отвечает определенное представление группы SU(6) : антисимметричной отвечает представление размерности 56, симметричной - представление размерности 20 и функции смешанной симметрии отвечает представление размерности 70.

Интересно следующее обстоятельство. Как хорошо известно, отклонение магнитных моментов ядер $He^3$ и $H^3$ от теоретических значений объясняется обменным магнитным моментом, который для $H^3$ положителен, а для $He^3$ отрицателен. В модели кварков протон есть аналог $He^3$ , а нейтрон - аналог $H^3$ и отклонение магнитных моментов протона $\mu_p$ = 2,79 и нейтрона $\mu_n$ = -1,91 от теоретических согласуется с тем, что имеет место для $He^3$ и $H^3$ .

Автор выражает искреннюю благодарность академику Н.Н. Боголюбову за предложенную задачу и внимание.

## Литература

Bottom part of back cover: